\documentclass[a4paper,10pt,twoside]{cpc-hepnp}

\usepackage{multicol}
\usepackage{graphicx}
\usepackage{booktabs}
\usepackage{amssymb,bm,mathrsfs,bbm,amscd}
\usepackage[tbtags]{amsmath}
\usepackage{lastpage}
\usepackage{CJK}
\usepackage{multirow}
\usepackage{color}

\begin{document}
\begin{CJK*}{GBK}{song}



\title{Inflation in de Sitter spacetime and CMB large scales anomaly}

\author{%
Dong Zhao$^{1:1)}$\email{zhaod@ihep.ac.cn}%
\quad Ming-Hua Li$^{2}$
\quad Ping Wang$^{1}$
\quad Zhe Chang$^{1}$
}
\maketitle

\address{
$^1$ Institute of High Energy Physics, Chinese Academy of Sciences, Beijing 100049, China\\
$^2$ School of Physics and Engineering, Sun Yat-Sen University, Guangzhou 510275, China\\
}

\begin{abstract}
The influence of cosmological constant type dark energy in the early universe is investigated. This is accommodated by a new dispersion relation in de Sitter spacetime. We perform a global fitting to explore the cosmological parameters space by using the CosmoMC package with the recently released Planck TT and WMAP Polarization datasets. Using the results from global fitting, we compute a new CMB temperature-temperature spectrum. The obtained TT spectrum has lower power compared with the one based on $\Lambda$CDM model at large scales.
\end{abstract}




\begin{multicols}{2}

\section{Introduction}
\label{intro}
The anisotropy of Cosmic Microwave Background radiation (CMBR) was first discovered by NASA's Cosmic Background Explorer (COBE) satellite in the 1990s\cite{cobe}. The results were later confirmed by the balloon experiments and the Wilkinson Microwave Anisotropy Probe (WMAP) satellite\cite{wmap9}. The observation data can be well fitted by the $\Lambda$CDM (cold dark matter plus dark energy in a form of the cosmological constant $\Lambda$) model. With the help of the Planck data \cite{planck2013}, one can now place an unprecedentedly precise constraint on six cosmological parameters (with an accuracy down to $10\%$ level)\cite{H2010-}.

However, for the power spectrum of CMBR, the observed values of $C_{\ell}$ for low $\ell$, especially for the quadrupole component $\ell =2$, are smaller than that predicted by the standard cosmological model. H. Liu and T.-P. Li \cite{Li2010-}\cite{Li2011-} proposed that the CMB quadrupole in the WMAP data is artificial and the corresponding values should actually be near zero. Contrarily, C. Bennett \textit{et al.} \cite{Bennett2011} reexamined the WMAP 7-year data carefully and reported that the quadrupole amplitude value is consistent with that predicted by the the $\Lambda$CDM model at a $95\%$ confidence level and shows no anomalies.

In a previous paper, we proposed inflation in a de Sitter spacetime that could possibly alleviate this controversy\cite{liminghua}. In fact, a de Sitter Universe is a cosmological solution of Einstein's field equations in general relativity with a positive cosmological constant $\Lambda$. We considered the possible effects of dark energy in form of a cosmological constant during the inflationary period. In this scenario, the cosmological constant type dark energy was once predominant in the early universe. The dynamics of the universe is accommodated by a new dispersion relation---the dispersion relation in de Sitter spacetime. We found that for certain cosmological parameter values, the modified inflation model gives a CMB TT power spectrum with lower power at large scales, alleviating the low-$\ell$ multipole issue.

In this paper, we would use the Markov Chain Monte Carlo sampler (CosmoMC)\cite{cosmomc} to explore the cosmological parameters space. The recently released Planck TT\cite{planck2013} and WMAP Polarization\cite{wmap9} datasets will be used in our global fitting. More stringent constraints are presented on the cosmological parameters. We obtain values of the cosmological parameters from global fitting and compute a new TT spectrum with the Code for Anisotropies in the Microwave Background (CAMB)\cite{camb}.

The rest of the paper is organized as follows. Section 2 is devoted to setup of the inflation in de Sitter spacetime. In section 3, we present a global fitting with combined datasets and obtain numerical results. Discussions and conclusions are given in section 4.

\section{Inflation in de Sitter Spacetime}
De Sitter spacetime is a vacuum solution of the Einstein's field equations with a positive cosmological constant. It can be realized as a four-dimensional pseudo-sphere imbedded in a five dimensional Minkowski flat space(with coordinates $\xi_{\mu}$, $\mu =0,1,2,3,4$) \cite{S2004-Cosmic}

\begin{equation}
\label{xi}
\begin{array}{l}
\displaystyle
-\xi_0^2 + \xi_1^2 + \xi_2^2 +\xi_3^2 +\xi_4^2=\frac{1}{K}=R^2 \ ,\\[5mm]
ds^2=-d\xi_0^2 +d\xi_1^2 +d\xi_2^2 +d\xi_3^2 +d\xi_4^2\ ,
\end{array}
\end{equation}
where $K$ is the Riemannian curvature and $R$ is the radius of de Sitter spacetime.

For a free particle with mass $m_0$, the five dimensional angular momentum $M_{\mu \nu}$ is defined as
\begin{equation}\label{angular}
M_{\mu \nu}\equiv m_0\left(\xi_{\mu}\frac{d\xi_{\nu}}{ds}-\xi_{\nu}\frac{d\xi_{\mu}}{ds}\right)\ ,~~~~~\mu,\nu =0,1,2,3,4\ ,
\end{equation}
where $s$ is the affine parameter. For the rest of the article, the Latin indices (i.e. $i,j,k$, etc.) run from 1 to 3 and the Greek indices (i.e. $\mu,\nu,\alpha,\beta$, etc.) run from 0 to 3.

The momentum of a free particle in de Sitter spacetime is defined as
\begin{equation}
\label{momentum}
P_\mu \equiv R^{-1}M_{4\mu}\ .
\end{equation}
The angular momentum $J_{\mu \nu}$ in de Sitter spacetime is defined as
\begin{equation}
\label{angular-mn} J_{\mu \nu}\equiv M_{\mu \nu} =
x_\mu P_\nu-x_\nu P_\mu\ .
\end{equation}

An invariant can be constructed in terms of the angular momentum $J^{\mu\nu}$,

\begin{equation}
\begin{array}{c}
\label{casimir}
m^2_0=\displaystyle\frac{K}{2}J_{\mu\nu}J^{\mu\nu}=E^2-{\bf P}^2+\frac{K}{2} J_{ij}J^{ij}~,\\
E=P_0~,~~~~{\bf P}=(P_1,~P_2,~P_3)~.
\end{array}
\end{equation}
The generators $\hat{M}_{\mu\nu}$ of de Sitter group $SO(1,4)$ are of the form
\begin{equation}
\label{L-operator}
\hat{M}_{\mu\nu}\equiv-i\left(\xi_{\mu}\frac{\partial}{\partial\xi^{\nu}}-\xi_{\nu}
\frac{\partial}{\partial\xi^{\mu}}\right)\ .
\end{equation}

The Klein-Gordon equation for a free scalar can be written as

\begin{equation}
\label{motion}
\left(\frac{K}{2}\hat{M}_{\mu\nu}\hat{M}^{\mu\nu}-m^2_0\right)\phi(\xi^{\mu})=0~.
\end{equation}
The equation (\ref{motion}) can be solved analytically \cite{S2004-Cosmic}. The dispersion relation for a free scalar in de Sitter spacetime is

\begin{equation}
\label{dispersion}
E^2=m_0^2+k^2+K(2n+l)(2n+l+2)\ ,
\end{equation}
where $n$ and $l$ respectively denotes the radial and the angular quantum number of the system. For massless particles like photons, the above dispersion relation becomes

\begin{equation}
\begin{array}{c}
\label{dispersion1}
\omega^2= k^2+\varepsilon^{*2}_{\gamma}\ ,\\ [0.5cm]
\varepsilon^{*}_{\gamma}\equiv \sqrt{K(2n_\gamma +l_\gamma)(2n_\gamma +l_\gamma +2)}\ ,
\end{array}
\end{equation}
where $w$ and $k$ are frequency and wavenumber of the photon.

Now, we calculate the primordial power spectrum ${\cal P}_{\delta \phi}(k)$ in de Sitter spacetime.
The inflation field is denoted by $\delta\phi({\bf x},t)$.
In the momentum representation, the evolution equation of the primordial perturbation is given by \cite{A2002-}

\begin{equation}
\delta\sigma^{\prime\prime}_{\bf k} +\left(\omega^2-\frac{2}{\tau^2}\right)\delta\sigma_{\bf k}=0\ ,
~~~~~\omega^2\equiv k^2+\varepsilon^{*2}_{\gamma}~,
\label{KGeq5}
\end{equation}
where
\begin{equation}
\delta\sigma_{\bf k} \equiv a\delta\phi_{\mathbf{k}}\ .
\label{sig}
\end{equation}
A prime represents differentiation with respect to the conformal time $\tau$ \cite{Dodelson2003-}.
The above equation has an exact particular solution

\begin{eqnarray}
\label{sigma}
\delta\sigma_{\bf k} &=& \frac{e^{-i\omega\tau}}{\sqrt{2\omega}}\left(
1+\frac{i}{\omega\tau}\right) \nonumber \\
&=& \frac{e^{-i\sqrt{k^2+\varepsilon^{*2}_{\gamma}}~\tau}}{\sqrt{2}\left(k^2+\varepsilon^{*2}_{\gamma}\right)^{1/4}}
\left(1+ \frac{i}{\sqrt{k^2+\varepsilon^{*2}_{\gamma}}~\tau}\right)\ .
\end{eqnarray}

The power spectrum ${\cal P}_{\delta \phi}(k)$ is defined as \cite{A2002-}

\begin{equation}
\label{powerdef}
{\cal P}_{\delta \phi}(k) \equiv \frac{k^3}{2\pi^2}\,|\frac{1}{a}\,\delta \phi_{\textbf{k}}|^2\ .
\end{equation}
Considering the super-horizon criterion, one has
\begin{equation}
\label{ktau}
-k\tau = \frac{k}{aH} \ll 1\ ,
\end{equation}
where $a=-1/H\tau$ ($H$ is the Hubble parameter) \cite{A2002-}.
By making use of (\ref{sigma}), (\ref{powerdef}) and (\ref{ktau}), we obtain the primordial power spectrum\cite{liminghua}

\begin{eqnarray}
\label{powerdeltaphi}
{\cal P}_{\delta \phi}(k)=\frac{H^2}{4\pi^2}\cdot \frac{k^3}{\left({k^2+\varepsilon^{*2}_{\gamma}}\right)^{3/2}}\ .
\end{eqnarray}

For perturbations on small scales, we obtain the usual scale-invariant primordial power spectrum

\begin{equation}
{\cal P}_{\delta \phi}(k)~\simeq~k^0.
\end{equation}
For large scales, we have

\begin{equation}
{\cal P}_{\delta \phi}(k) \simeq \frac{H^2}{4\pi^2}\cdot k^3.
\end{equation}

The power spectrum of the comoving curvature perturbation $\mathcal{R}$ is usually parameterized as \cite{Dodelson2003-}

\begin{equation}
\label{powerdef2}
\mathcal{P}_{\mathcal{R}}(k)  = \frac{H^2}{\dot{\phi}^2}~{\cal P}_{\delta \phi}(k) \equiv A^2_s \left(\frac{k}{k_p} \right)^{n_s-1}\cdot \frac{(k/k_p)^3}{[{\left(k/k_p\right)^2+\varepsilon^{*2}}]^{3/2}}\  ,
\end{equation}
where $k_p=0.05~\rm{Mpc^{-1}}$ and $\varepsilon^*=\varepsilon^*_{\gamma}/k_p$. $A_s$ is the power amplitude and $n_s$ is the scalar spectral index.

We plot the primordial spectrum (\ref{powerdef2}) in Figure \ref{fig1}.
\begin{center}
\includegraphics[width=9cm]{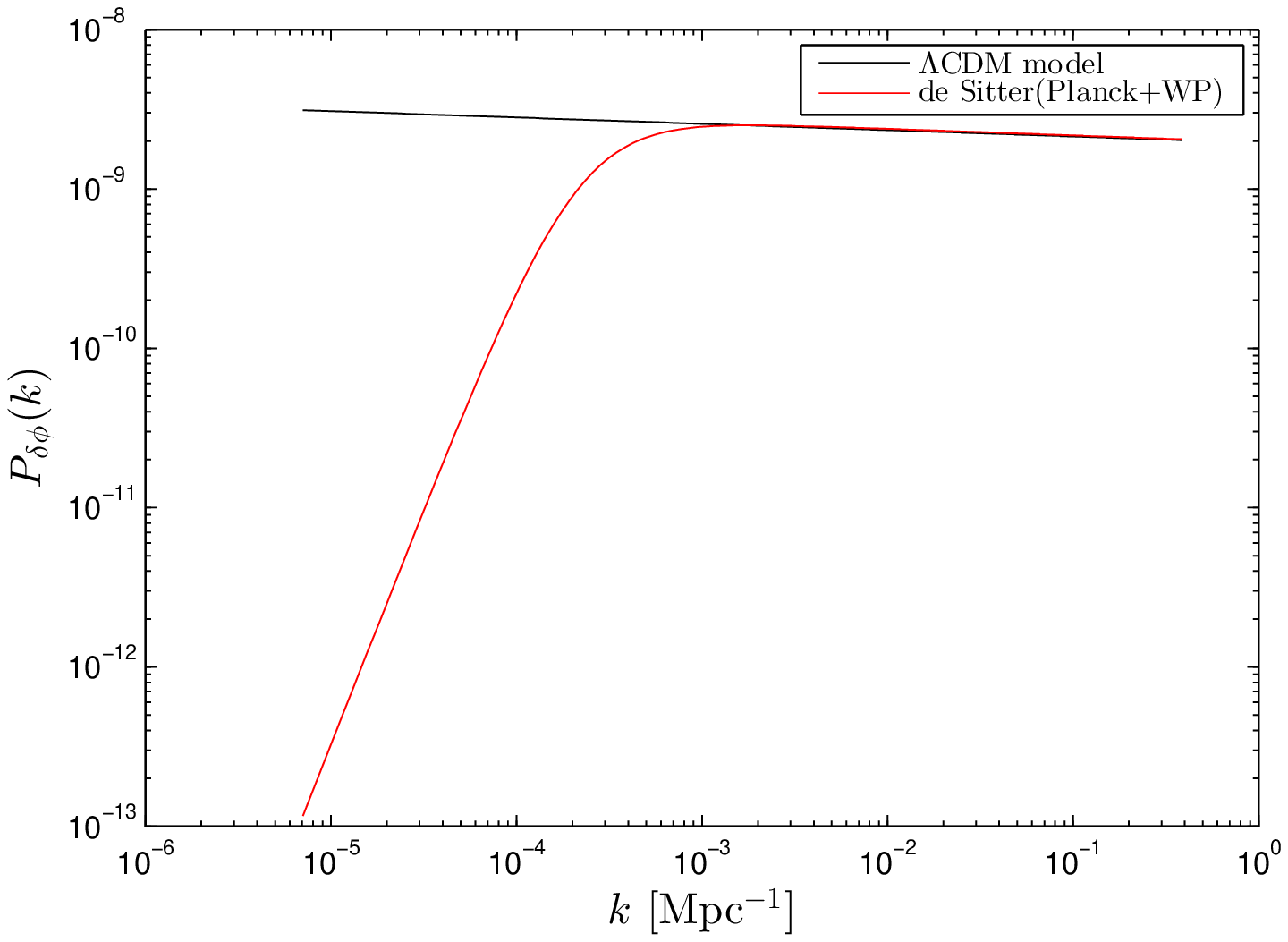}
\figcaption{\label{fig1}   The primordial power spectrum $P_{\delta\phi}(k)$. The black curve stands for the $P_{\delta\phi}(k)$ in standard $\Lambda$CDM model and the red curve represents the $P_{\delta\phi}(k)$ in de Sitter spacetime. One can see that there is a cutoff at large scales. It should lead to a suppressed TT spectrum when $\ell$ is small.}
\end{center}

\section{The Numerical Results}
\label{sec:3}
We use the Markov Chain Monte Carlo sampler (CosmoMC) to perform a global fitting of cosmological parameters. The Planck TT\cite{planck2013} and WMAP polarization\cite{wmap9} datasets will be used in the global fitting. There are six parameters originated in the $\Lambda$CDM model and one extra parameter $\varepsilon^*$. The results are shown in Table 1 and Figure 1.

\begin{center}
\footnotesize
\begin{tabular*}{80mm}{c@{\extracolsep{\fill}}ccc}
\hline
\toprule Parameter & $\Lambda$CDM model & de Sitter \\
\hline
$\Omega_bh^2$ & 0.02205$\pm$0.00028 & 0.02203$\pm$0.00028  \\

$\Omega_ch^2$ & 0.1199$\pm$0.0027 & 0.1200$\pm$0.0026 \\

$100\theta_{MC}$ & 1.04131$\pm$0.00063 & 1.04121$\pm$0.00063  \\

$\tau$ & $0.089_{-0.014}^{+0.012}$ & 0.098$\pm$0.015  \\

$n_s$ & 0.9603$\pm$0.0073 & 0.9585$\pm$0.0073  \\

$ln(10^{10}A^s)$ & $3.089_{-0.027}^{+0.024}$ & 3.106$\pm$0.030  \\


$100\varepsilon^*$ & - & 0.4266$\pm$0.1945  \\
\bottomrule
\end{tabular*}
\tabcaption{ \label{tab1}   The 68$\%$ limits for the cosmological parameters originated in the $\Lambda$CDM and de Sitter models with data combination Planck+WP.}
\end{center}

\end{multicols}

\begin{center}
\includegraphics[width=17cm]{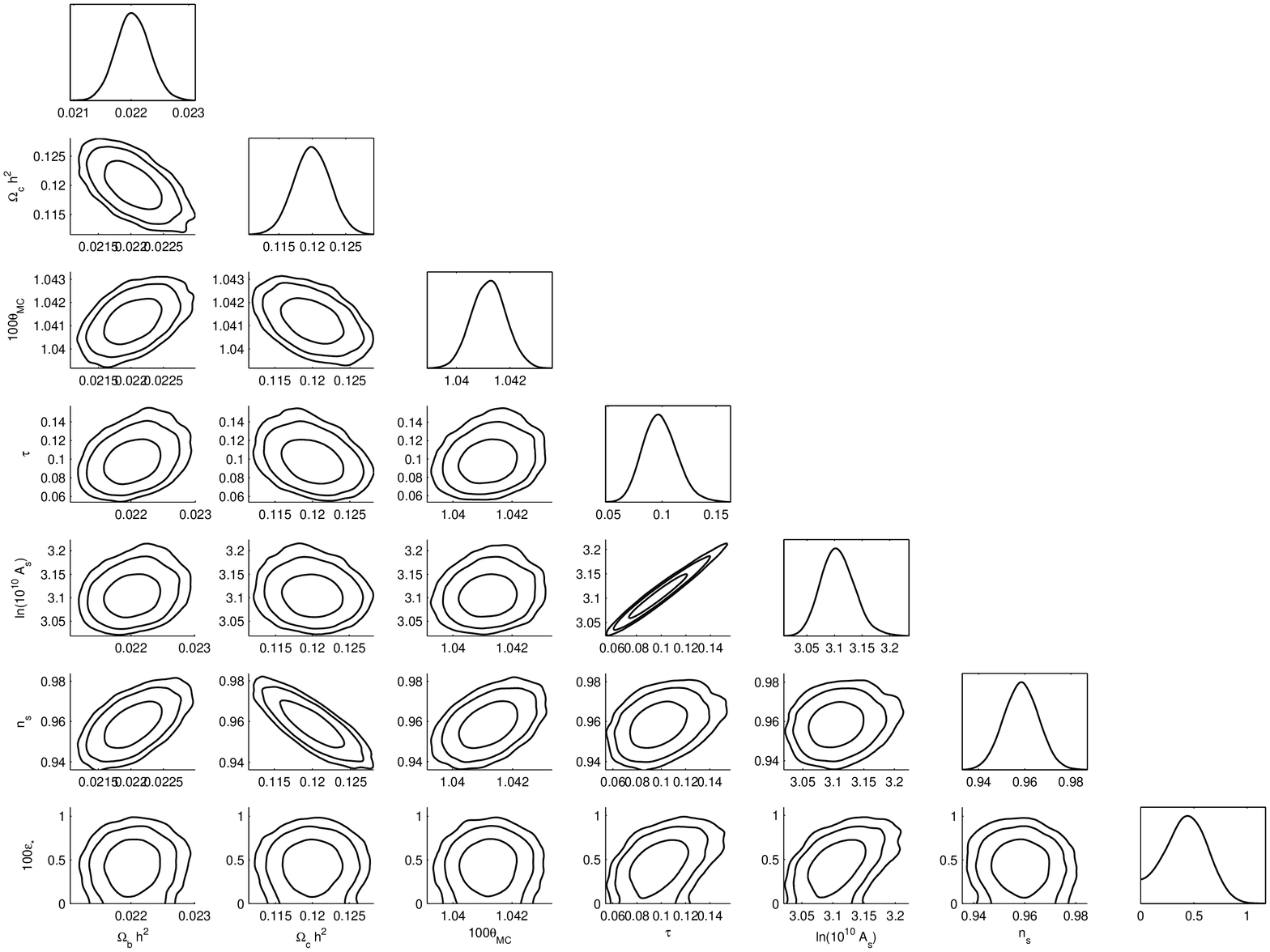}
\figcaption{\label{fig2} Contour plots and the likelihood distributions of the seven cosmological parameters with data combination Planck+WP in de Sitter spacetime.}
\end{center}

\begin{multicols}{2}
In Table \ref{tab1}, one can see that the values of the six basic parameters in de Sitter spacetime are similar with the ones based on standard cosmological model\cite{planck2013}. We are more interested in the new parameter $\varepsilon^*_{\gamma}$. With $\varepsilon^*_{\gamma}=2.13\times10^{-4}~\rm{Mpc^{-1}}$, we can calculate the wavelength related to $\varepsilon^*_{\gamma}$, i.e. $\lambda=4.69\times10^3~\rm{Mpc}$. The wavelength is comparable to the hubble horizon. From Figure \ref{fig2}, one can see that the combined datasets favour the new parameter $\varepsilon^*_{\gamma}$ is not zero at around 1$\sigma$ confidence level.

Finally, we use the modified version of CAMB to compute the new CMB temperature-temperature spectrum in de Sitter spacetime with the global fitting results. The obtained TT spectrum has been suppressed obviously at large scales.

\begin{flushleft}
\includegraphics[width=9cm]{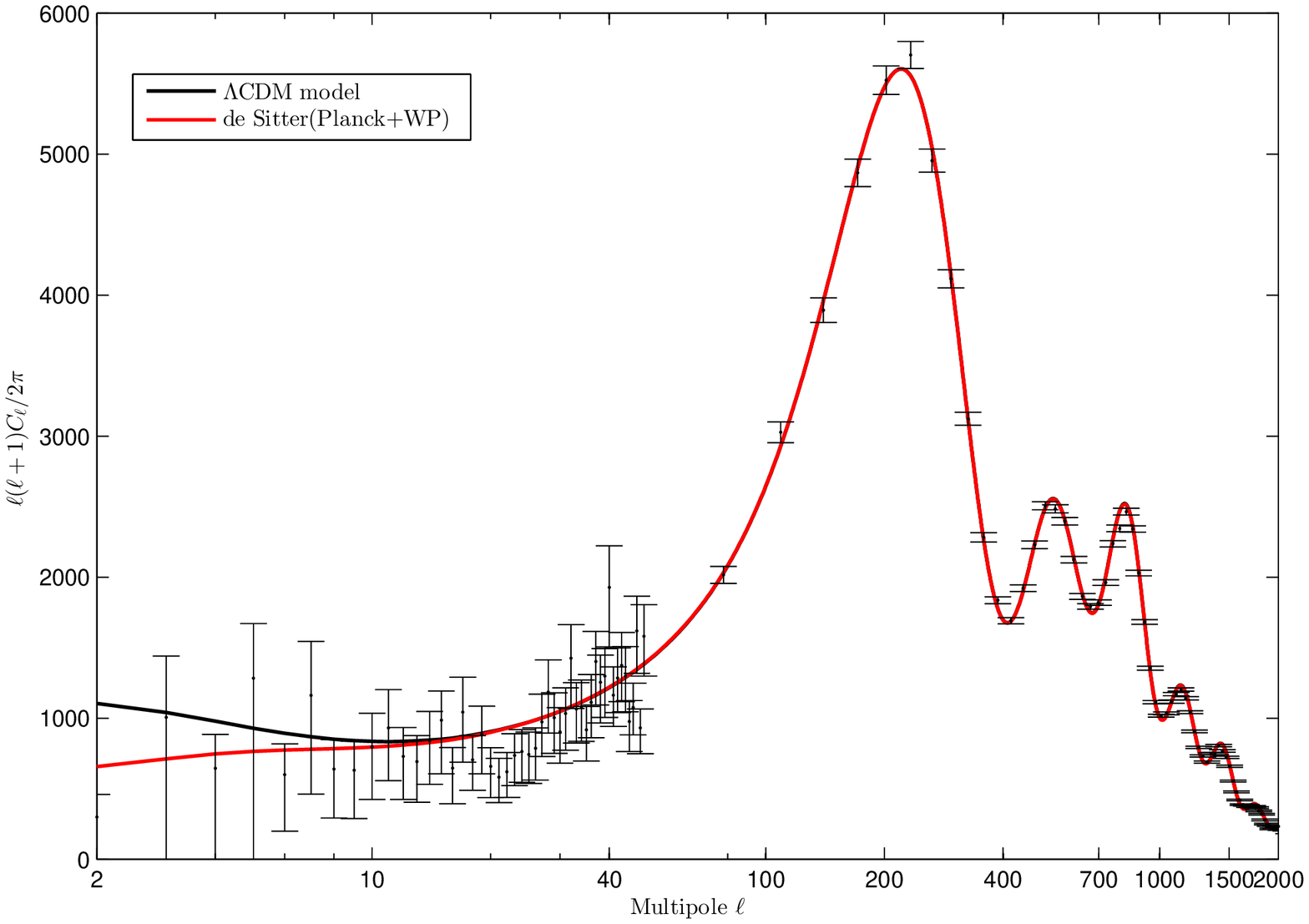}
\figcaption{\label{fig3} The TT spectra with data combination Planck+WP. The black curve represents TT spectrum based on the standard $\Lambda$CDM model and the red one indicates TT spectrum in de Sitter spacetime.}
\end{flushleft}

In Figure \ref{fig3}, the black curve indicates the best-fitting CMB TT spectrum in standard $\Lambda$CDM model and the red curve represents the TT spectrum in de Sitter spacetime. Compared to the TT spectrum in the $\Lambda$CDM model, the spectrum in de Sitter spacetime has been suppressed obviously when $\ell$ $<$ 20. Especially at quadrupole, the spectrum drops to the lowest point, which is almost half of the standard TT spectrum.

\section{Conclusions}

In this paper, we analyzed an inflation model in de Sitter spacetime and got modified primordial spectrum. Possible effects of cosmological constant type dark energy were considered during the inflationary period. We got a lower TT spectrum at large scales which refers to low-$\ell$ multipole anomaly. The new scenario has a new parameter, namely $\varepsilon^*_{\gamma}$. To constrain the cosmological parameters, we used the CosmoMC package to perform a global fitting with Planck TT and WMAP Polarization datasets. We found that the results of global fitting in de Sitter spacetime are similar to ones based on standard $\Lambda$CDM model. With $\varepsilon^*_{\gamma}=2.13\times10^{-4}~\rm{Mpc^{-1}}$, we calculated the wavelength related to $\varepsilon^*_{\gamma}$, i.e. $\lambda=4.69\times10^3~\rm{Mpc}$. The wavelength is comparable to the hubble horizon. The combined datasets favour that the new parameter $\varepsilon^*_{\gamma}$ is not zero at about 1$\sigma$ confidence level. The new TT spectrum shows lower energy at large scales, which just as we expected.

We are grateful to Dr. Hai-Nan Lin, Si-Yu Li, Sai Wang and Yu Sang for useful discussion. This work has been funded by the National Natural Science Fund of China under Grant no. 11375203.
\\

\end{multicols}

\clearpage

\end{CJK*}
\end{document}